\RequirePackage{ifpdf}
\ifpdf 
\documentclass[pdftex]{sigma}
\else
\documentclass{sigma}
\fi

\begin{document}

\allowdisplaybreaks

\renewcommand{\PaperNumber}{094}

\FirstPageHeading

\renewcommand{\thefootnote}{$\star$}

\ShortArticleName{Noncommutative Geometry: Fuzzy Spaces, the
Groenewold--Moyal Plane}

\ArticleName{Noncommutative Geometry: Fuzzy Spaces,\\ the
Groenewold--Moyal Plane\footnote{This paper is a contribution to
the Proceedings of the O'Raifeartaigh Symposium on
Non-Perturbative and Symmetry Methods in Field Theory (June
22--24, 2006, Budapest, Hungary). The full collection is available
at
\href{http://www.emis.de/journals/SIGMA/LOR2006.html}{http://www.emis.de/journals/SIGMA/LOR2006.html}}}

\Author{Aiyalam P. BALACHANDRAN and Babar Ahmed QURESHI}

\AuthorNameForHeading{A.P. Balachandran and B.A. Qureshi}
\Address{Department of Physics, Syracuse University, Syracuse, NY, USA} 

\Email{\href{mailto:bal@phy.syr.edu}{bal@phy.syr.edu}, \href{mailto:bqureshi@phy.syr.edu}{bqureshi@phy.syr.edu}}

\ArticleDates{Received September 22, 2006, in f\/inal form
December 14, 2006; Published online December 29, 2006}

\Abstract{In this talk, we review the basics concepts of fuzzy
physics and quantum f\/ield theory on the Groenewold--Moyal Plane
as examples of noncommutative spaces in physics. We introduce the
basic ideas, and discuss some important results in these f\/ields.
At the end we outline some recent developments in the f\/ield.}


\Classification{81R60; 46L65}

\section{Introduction}

Noncommutative geometry is a branch of mathematics due to
Gel'fand, Naimark, Connes, Rief\/fel and many
others~\cite{Connes,Connes1,Connes2}. Physicists in a very short
time adopted it and nowadays use this phrase whenever spacetime
algebra is noncommutative.

There are two such particularly active f\/ields in physics at
present
\begin{enumerate}\itemsep=0pt
\item Fuzzy Physics, \item Quantum Field Theory~(QFT) on the
Groenewald--Moyal Plane.
\end{enumerate}

Item 1 is evolving into a tool to regulate QFT's, and for
numerical work. It is an alternative to lattice methods. Item 2 is
more a probe of Planck-scale physics. This introductory talk will
discuss both items 1 and 2.

\section{History}

 The Groenewold--Moyal (G-M) plane is associated with noncommutative spacetime
 coordinates:
 \begin{gather*}
 [x_\mu,x_\nu]=i\theta_{\mu\nu}.
 \end{gather*}
It is an example where spacetime coordinates do not commute.

  The idea that spatial coordinates may not commute f\/irst occurs in a letter from Heisenberg to
Peierls~\cite{Pauli,Pauli1}. Heisenberg suggested that an
uncertainty principle such as
\begin{gather}
\Delta x_\mu\Delta x_\nu\geq\frac{1}{2}|\theta_{\mu\nu}|,\qquad
\theta_{\mu\nu}=\textrm{const}\nonumber
\end{gather}
can provide a short distance cut-of\/f and regulate quantum
f\/ield theories (qft's). In this letter, he apparently complains
about his lack of mathematical skills to study this possibility.
Peierls communicated this idea to Pauli, Pauli to Oppenheimer and
f\/inally Oppenheimer to Snyder. Snyder wrote the f\/irst paper on
the subject~\cite{Snyder}. This was followed by a paper of
Yang~\cite{Yang}.

In mid-90's, Doplicher, Fredenhagen and
Roberts~\cite{Doplicher,Doplicher1} systematically constructed
unitary quantum f\/ield theories on the G-M plane and its
generalizations, even with time-space noncommutativity.

Later string physics encountered these structures.

\renewcommand{\thefootnote}{\arabic{footnote}}
\setcounter{footnote}{0}

\section{What is noncommutative geometry}
According to Connes~\cite{Connes,Connes1,Connes2}, noncommutative
geometry is a spectral triple,
\[
(\mathcal{A}, D, \mathcal{H}),
\]
where $\mathcal{A}$\ =\ a $C^{\ast}$-algebra, possibly
noncommutative, $D$\ =\ a Dirac operator, $\mathcal{H}\ =\ $a
Hilbert space on which they are represented.

If $\mathcal{A}$ is a commutative $C^\ast$-algebra, we can recover
a Hausdorf\/f topological space  on which $\mathcal{A}$ are
functions,using theorems of Gel'fand and Naimark. But that is not
possible if $\mathcal{A}$ is not commutative. But it is still
possible to formulate qft's using the spectral triple. A class of
examples of noncommutative geometry with $\mathcal{A}$
noncommutative is due to Connes and Landi~\cite{Landi}.

If some of the strict axioms are not enforced then the examples
include $SU(2)_q$, fuzzy spaces, the G-M plane, and many more.

The introduction of noncommutative geometry has introduced a
conceptual revolution. Mani\-folds are being replaced by their
``duals'', algebras, and these duals are being ``quantized'', much
as in the passage from classical to quantum mechanics.

\section{Fuzzy physics}
In what follows, we sketch the contents of ``fuzzy physics''.
Reference~\cite{Bal1} contains a detailed survey. For pioneering
work on fuzzy physics, see~\cite{hoppe,madore,madore1}.

\subsection{What is fuzzy physics~\cite{Bal1}}
We explain the basic ideas of fuzzy physics by a two-dimensional
example: $S_F^2$.

Consider the two-sphere $S^2$. We quantize it to regularize by
introducing a short distance cut-of\/f. For example in classical
mechanics, the number of states in a phase space volume
\[
\Delta V  =  d^3p  d^3q
\]
is inf\/inite. But we know since Planck and Bose that on
quantization, it becomes
\[ \frac{\Delta V}{h^3} = \textrm{f\/inite}.
\]
This is the idea behind fuzzy regularization.

In detail, this regularization works as follows on $S^2$. We have
\begin{gather*}
S^2=[ {\vec x} \in\mathbb{R}^3:\ {\vec x}\cdot{\vec x} =
r^2].
\end{gather*}
Now consider angular momentum $L_i$:
\begin{gather*}
[L_i,L_j]=i\epsilon_{ijk} L_k,\qquad {\vec L}^2=l(l+1).
\end{gather*}
 Set
\begin{gather*}
\hat x_i = r \frac{L_i}{\sqrt{l(l+1)}}\quad \Rightarrow
\\
\hat x\cdot \hat x=r^2,\qquad [\hat x_i,\hat
x_j]=\frac{r}{\sqrt{l(l+1)}}i\epsilon_{ijk}\hat x_k,
\end{gather*}
where $\hat x_i \in {\rm Mat}_{2l+1}\equiv \textrm{space of
}(2l+1)\times (2l+1)$ matrices. As $l \to \infty$, they become
commutative. They give the fuzzy sphere $S^2_F$ of radius $r$ and
dimensions $2l+1$.

\subsection{Why is this space fuzzy}
 As $\hat x_i$, $\hat x_j$ $(i \neq j)$ do not
commute, we cannot sharply localize $\hat x_i$. Roughly in a
volume $4\pi r^2$ there are $(2l+1)$ states.

\subsection{Field theory on fuzzy sphere}
A scalar f\/ield on fuzzy sphere is def\/ined as a polynomial in
$\hat x_i$, i.e.,
\[
\textrm{A scalar f\/ield}\ \Phi\ =\ \textrm{A polynomial in}\ \hat
x_i\  =\ \textrm{A }(2l+1)-\textrm{dimensional matrix.}
\]

Dif\/ferentiation is given by inf\/initesimal rotation:
\[
\mathcal{L}_i \Phi = [L_i,\Phi].
\]
A simple rotationally invariant scalar f\/ield action is given by
\begin{gather*}
S(\Phi) = \mu\,\mathrm{Tr}\,[L_i,\Phi]^\dag[L_i,\Phi] +
\frac{m}{2}\,\mathrm{Tr}\,(\Phi^\dag\Phi) +
\lambda\,\mathrm{Tr}\,(\Phi^\dag\Phi)^2.
\end{gather*}
Simulations have been performed~\cite{Xavier,Xavier1} on the
partition function $\mathcal{Z}=\int d\Phi e^{-S(\Phi)}$ of this
model and the major f\/indings include the following:
\begin{itemize}\itemsep=0pt
 \item Continuum limit exists.
 \item If
 \[
  \Phi = \sum c_{lm} \hat Y_{lm},\qquad  \hat Y_{lm}=\textrm{spherical tensor},
  \]
  then there are three phases:
  \begin{enumerate}
   \item Disordered :      $\langle \sum |c_{lm}|^2 \rangle = 0$.
   \item Uniform ordered:     $\langle |c_{00}|^2 \rangle \neq 0$,
   $\langle |c_{lm}|^2 \rangle = 0$ for $l\neq 0.$
   \item Non-uniform ordered:  $\langle |c_{1m}|^2 \rangle \neq 0$,
   $\langle |c_{lm}^2 \rangle = 0$ for $l\neq 1$.
  \end{enumerate}

 The last one is an analogue of the Gupser--Sondhi phase~\cite{Sondhi,Sondhi1,Sondhi2}.
\end{itemize}

\subsubsection{Dirac operator}
$S^2_F$ has a Dirac operator including instantons and with no
fermion
doubling~\cite{GKP1,GKP1+,giorgio,watamura,monopole,fermion,fermion1}.

Also $S^2_F$ can nicely describe topological features. Hence it
seems better suited for preserving symmetries than lattice
approximations.

\subsubsection{Supersymmetry}
If we replace $SU(2)$ by $OSp(2,1)$, the fuzzy sphere becomes the
$N=1$ supersymmetric fuzzy sphere and can be used to simulate
supersymmetry~\cite{GKP1,GKP1+,fuzzyS,klimcik1,klimcik1+,seckin1,susybreaking,seckin2}.
Simulations in this regard are already starting.

\subsubsection{Strings~\cite{Szabo}}
If $N$ $D$-branes are close, the transverse coordinates $\Phi_i$
become $N \times N$ matrices with the action given by
\begin{gather*}
\mathrm{S} = \lambda \mathrm{Tr}\,[\Phi_i,\Phi_j]^\dag
[\Phi_i,\Phi_j] + i f_{ijk} \Phi_i\Phi_j\Phi_k,
\end{gather*}
where $f_{ijk}$ are totally antisymmetric.

The equations of motion
\begin{gather*}
[\Phi_i,\Phi_j] \cong i f_{ijk}\Phi_k
\end{gather*}
 give solutions when $f_{ijk}$ are structure constants of a simple compact Lie group. Thus we can have
\[
\Phi_i=cL_i,\qquad f_{ijk}=c\epsilon_{ijk},\qquad
c=\textrm{const},\qquad
 L_i=\textrm{angular momentum operators.}
 \]
If $L_i$ form an irreducible set, then we have
\[
{\vec L}\cdot{\vec L} = l(l+1),\qquad (2l+1)=N,
\]
and we have one fuzzy sphere. Or we can have a direct sum of
irreducible representations:
\[
L_i = \oplus L_i^{l_k},\qquad {\vec L}^{l_k}\cdot{\vec
L}^{l_k}=l_k(l_k+1),\qquad \sum (2l_k+1)=N.
\]
Then we have many fuzzy spheres.

Stability analysis of these solutions including numerical studies
has been done by many groups.

\section{The G-M Plane}

\subsection{Quantum gravity and spacetime noncommutativity: heuristics}
The following arguments were described by Doplicher, Fredenhagen
and Robert in their work in support of the necessity of
noncommutative spacetime at Planck scale.

\subsubsection{Space-space noncommutativity}
In order to probe physics at the Planck scale $L$, the Compton
wavelength $\frac{\hbar}{Mc}$ of the probe must fulf\/ill
\begin{gather*}
\frac{\hbar}{Mc} \leq  L\qquad {\rm  or}\qquad M \geq
\frac{\hbar}{Lc} \simeq \textrm{Planck mass}.
\end{gather*}
Such high mass in the small volume $L^3$ will strongly af\/fect
gravity and can cause black holes to form. This suggests  a
fundamental length limiting spatial localization.

\subsubsection{Time-space noncommutativity}
Similar arguments can be made about time localization. Observation
of very short time scales requires very high energies. They can
produce black holes and black hole horizons will then limit
spatial resolution suggesting
\begin{gather*}
\Delta t \Delta |{\vec x}|  \geq L^2,\qquad L = \textrm{a
fundamental length.}
\end{gather*}
The G-M plane {\it models} above spacetime uncertainties.

\subsection{What is the G-M plane}
The Groenewald--Moyal plane $\mathcal{A}_\theta(\mathbb{R}^{d+1})$
consists of functions $\alpha,\beta,\dots$ on $\mathbb{R}^{d+1}$
with the $\ast$-pro\-duct
\begin{gather*}
\alpha \ast \beta = \alpha e^{\frac{i}{2} {\overleftarrow
\partial}_\mu \theta^{\mu\nu} {\overrightarrow
\partial}_\nu} \beta.
\end{gather*}
For spacetime coordinates, this implies,
\begin{gather*}
[x_\mu,x_\nu]_\ast= x_\mu\ast x_\nu-x_\nu\ast x_\mu=
i\theta_{\mu\nu}.
\end{gather*}
Conversely these coordinate commutators imply the general
$\ast$-product up to certain equivalencies.

The G-M plane also emerges in quantum Hall ef\/fect and string
physics.

\subsection[How the G-M plane emerges from quantum Hall effect and strings]{How
the G-M plane emerges from quantum Hall ef\/fect and strings}

\subsubsection[Quantum Hall effect(the Landau problem)~\cite{ezawa}]{Quantum
Hall ef\/fect (the Landau problem)~\cite{ezawa}}

Consider an electron in 1--2 plane and an external magnetic
f\/ield ${\vec B}=(0,0,B)$ perpendicular to the plane. Then the
Lagrangian for the system is
\begin{gather*}
L = \frac{1}{2}m\dot x_a^2 + e\dot x_a^2 A_a,
\end{gather*}
where
\begin{gather*}
A_a = -\frac{B}{2}\epsilon_{ab}x^b,\qquad a,b=1,2,
\end{gather*}
is the electromagnetic potential and $x_a$ are the coordinates of
the electron.

Now if $eB  \to \infty$, then
\begin{gather*}
L\ \sim\ \frac{eB}{2} (\dot x_1 x_2 - \dot x_2 x_1).
\end{gather*}
This means that on quantization we will have
\begin{gather*}
[\hat x_a , \hat x_b] = \frac{i}{eB} \epsilon_{ab}
\end{gather*}
which def\/ines a G-M plane.

\subsubsection{Strings~\cite{witten}}
Consider open strings ending on $Dp$-Branes. If there is a
background two-form Neveu--Schwarz f\/ield given by the constants
$B_{ij}=-B_{ji}$, then the action is given by
\begin{gather*}
S_\Sigma  =  \frac{1}{4\pi\alpha'} \int_{\Sigma} \big[g_{ij}
\partial_a x^i \partial_a x^j - 2\pi\alpha' B_{ij} \partial_a x^i
\partial_b x^j \epsilon^{ab}
 +\textrm{spinor terms}\big]d\sigma dt.
\end{gather*}
As $B\to \infty$ or equivalently $g_{ij}\to 0$,
\begin{gather*}
S_\Sigma  =  -\frac{2\pi e}{4\pi}  \int_{\Sigma} B_{ij} dx^i\wedge
dx^j  =  \left[\int_{\partial\Sigma^0} -
\int_{\partial\Sigma^1}\right]  eB_{ij} x^i \frac{dx^j}{dt}\\
\Rightarrow\quad e [B_{ij} \hat x^j , \hat x^k]  =
i\delta_{ik}\qquad \textrm{or} \qquad
 [\hat x_j,\hat x_k]  =  \frac{i}{e}(B^{-1})_{jk}
\end{gather*}
which is just a G-M plane.

\begin{figure}[t]
\centerline{\includegraphics[width=8cm]{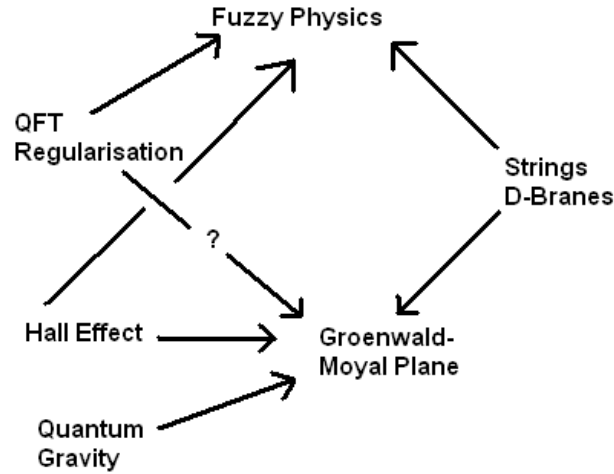}} \caption{The
tangled web:  emergence of noncommutative spaces from dif\/ferent
f\/ields.\label{f1}}
\end{figure}

Fig.~\ref{f1} indicates  dif\/ferent sources wherefrom fuzzy
physics and the G-M plane emerge. The question mark is to indicate
that the G-M plane may not regularize qft's.

\subsection{Prehistory (before 2004/2005)}

Until 2004/2005, much work was done on
\begin{itemize}\itemsep=0pt
\item QFT's on the G-M plane and its renormalization theory,
uncovering the phenomenon of UV/IR mixing~\cite{shiraz}. \item
Phenomenology, including the study of the ef\/fects of
noncommutativity on Lorentz invariance violation (from
$\theta_{\mu\nu}$ in $[x_\mu,x_\nu]_\ast=i\theta_{\mu\nu}$), C, CP
and CPT.
\end{itemize}

\subsection{Modern era}
In 2004/2005, Chaichian et al.~\cite{Chaichian,Chaichian1} and
Aschieri et al.~\cite{Aschieri,Aschieri1} applied the Drinfel'd
twist~\cite{Drinfeld} which restores full dif\/feomorphism
invariance (with a twist in the ``coproduct'') despite the
presence of constants $\theta_{\mu\nu}$ in $[\hat x_\mu,\hat
x_\nu]=i\theta_{\mu\nu}$. This twist also twists
statistics~\cite{Bal2,Bal2+}\footnote{There are claims to the
contrary, see~\cite{anca, zahn, bald} for the debate.}.

Much of this was known to Majid~\cite{Majid}, Oeckl~\cite{Oeckl},
Fiore and Schupp~\cite{Fiore,Fiore1,Fiore2} and
Watts~\cite{Watts,Watts1}. So the Drinfel'd twist twists both
\begin{enumerate}\itemsep=0pt
\item action of dif\/feomorphisms, and \item exchange statistics.
\end{enumerate}

This brings into question much of the prehistory-analysis.
Examples include the following new results:
\begin{enumerate}\itemsep=0pt
\item The Pauli principle can be violated on the G-M plane. \item
(Twisted) Lorentz invariance need not be violated even if
$\theta_{\mu\nu}\neq 0$. \item There need be no
ultraviolet-infrared (UV-IR) mixing in the absence of gauge
f\/ields~\cite{Bal3}.
\end{enumerate}

There is also a striking, clean separation of matter from gauge
f\/ields due to the Drinfel'd twist~\cite{Bal4}, (in the sense
that they have to be treated dif\/ferently) reminiscent of the
distinction between particles and waves in the classical theory.

Literature should be consulted for details of these developments.

\subsection*{Acknowledgments}
This work was supported by DOE under grant number
DE-FG02-85ER40231.

\LastPageEnding


\begin{thebibliography}{99}
\footnotesize
\bibitem{Connes}
Connes A., Noncommutative geometry, San Diego, CA, Academic Press,
1994.

\bibitem{Connes1} Varilly J.C., Figueroa H., Gracia-Bondia J.M., Elements of noncommutative geometry, Boston, Birkhauser,
 2000.

\bibitem{Connes2} Landi  G., Introduction to noncommutative spaces
and their geometries, New York, Springer Verlag, 1997.


\bibitem{Pauli}
Jackiw R., Physical instances of noncommuting coordinates, {\it
Nuclear Phys. Proc. Suppl.}, 2002, V.108, 30--36,
\href{http://arxiv.org/abs/hep-th/0110057}{hep-th/0110057}.

\bibitem{Pauli1} Pauli W.,  Letter of Heisenberg to Peirels (1930),
in Wolfgang Pauli, Scientific Correspondence,
Vol.~II,  Springer Verlag, 1985, 15--15.


\bibitem{Snyder}
 Snyder H., Quantized space-time, {\it Phys. Rev.}, 1947, V.71, 38--41.

\bibitem{Yang}
 Yang C.N., On quantized space-time, {\it Phys. Rev.}, 1947, V.72, 874--874.

\bibitem{Doplicher}
 Doplicher S., Fredenhagen K., Roberts J., Space-time quantization
induced by classical gravity, {\it Phys. Lett.~B}, 1994, V.331,
39--44.

\bibitem{Doplicher1} Doplicher S., Fredenhagen K., Roberts J., The quantum
structure of space-time at the Planck scale and quantum f\/ields,
{\it Comm. Math. Phys.}, 1995, V.172, 187--220,
\href{http://arxiv.org/abs/hep-th/0303037}{hep-th/0303037}.

\bibitem{Landi}
 Connes A., Landi G., Noncommutative manifolds: the instanton
algebra and isospectral deformations, {\it Comm. Math. Phys.},
2001, V.221, 141--159,
\href{http://arxiv.org/abs/math.QA/0011194}{math.QA/0011194}.

\bibitem{Bal1}
Balachandran A.P., Kurkcuoglu S., Vaidya S.,  Lectures on fuzzy
and fuzzy susy physics, World Scientif\/ic, to appear,
\href{http://arxiv.org/abs/hep-th/0511114}{hep-th/0511114}.

\bibitem{hoppe}
 Hoppe J., Quantum theory of a massless relativistic surface and a
two dimensional bound state problem, PhD Thesis, MIT, 1982.

\bibitem{madore}
Madore J., The fuzzy sphere, {\it Classical Quantum Gravity},
1992, V.9, 69--88.


\bibitem{madore1} Madore J., An introduction to non-commutative dif\/ferential
geometry and its physical applications, Cambridge, Cambridge
University Press, 1995.

\bibitem{Xavier}
 Martin X., A matrix phase for the $\phi^4$ scalar f\/ield on the fuzzy
sphere, {\it JHEP}, 2004, N~4, Paper~077, 15~pages,
\href{http://arxiv.org/abs/hep-th/0402230}{hep-th/0402230}.

\bibitem{Xavier1} Flores F.G., O'Connor D., Martin X., Simulating the scalar f\/ield on
the fuzzy sphere, in Proceedings for the XXIIIrd International
Symposium on Lattice Field Theory, {\it PoSLat2005}, 2006, 262, 6
pages,
\href{http://arxiv.org/abs/hep-lat/0601012}{\mbox{hep-lat/0601012}}.

\bibitem{Sondhi}
 Gubser S.S.,  Sondhi S.L., Phase structure of non-commutative
scalar f\/ield theories, {\it Nuclear Phys. B}, 2001, V.605,
395--424,
\href{http://arxiv.org/abs/hep-th/0006119}{hep-th/0006119}.

\bibitem{Sondhi1} Ambjorn J., Catterall S., Stripes from
(noncommutative) stars, {\it Phys. Lett. B}, 2002, V.549,
253--259,
\href{http://arxiv.org/abs/hep-lat/0209106}{\mbox{hep-lat/0209106}}.

\bibitem{Sondhi2} Medina J., Bietenholz W., Hofheinz F.,
 O'Connor D., Field theory simulations on a fuzzy sphere -- an
alternative to the lattice, {\it PoSLAT2005}, 2006,  263, 6 pages,
\href{http://arxiv.org/abs/hep-lat/0509162}{hep-lat/0509162}.



\bibitem{GKP1} Grosse H., Klim\v{c}ik C., Pre\v{s}najder P., Field theory on a supersymmetric
lattice,  {\it Comm. Math. Phys.}, 1997, V.185, 155--175,
\href{http://arxiv.org/abs/hep-th/9507074}{hep-th/9507074}.

\bibitem{GKP1+} Grosse H.,
Klim\v{c}ik C., Pre\v{s}najder P., $N=2$ superalgebra and
non-commutative geometry,
\href{http://arxiv.org/abs/hep-th/9603071}{hep-th/9603071}.


\bibitem{giorgio}
Balachandran A.P., Immirzi G., The fuzzy Ginsparg--Wilson algebra:
a solution of the fermion doubling problem, {\it Phys. Rev.~D},
2003, V.68, 065023, 7 pages,
\href{http://arxiv.org/abs/hep-th/0301242}{hep-th/0301242}.



\bibitem{watamura} Carow-Watamura U., Watamura S., Chirality and Dirac operator on noncommutative
sphere, {\it Comm. Math. Phys.}, 1997, V.183, 365--382,
\href{http://arxiv.org/abs/hep-th/9605003}{hep-th/9605003}.

\bibitem{monopole}
Baez S., Balachandran A.P., Ydri B., Vaidya S.,
 Monopoles and solitons in fuzzy physics, {\it Comm. Math. Phys.}, 2000, V.208,
787--798,
\href{http://arxiv.org/abs/hep-th/9811169}{hep-th/9811169}.

\bibitem{fermion}
Balachandran A.P., Govindarajan T.R., Ydri B., The fermion doubling problem and noncommutative geo\-met\-ry, 
{\it Modern Phys. Lett.~A}, 2000, V.15,
1279--1286,
\href{http://arxiv.org/abs/hep-th/9911087}{hep-th/9911087}.

\bibitem{fermion1}Balachandran A.P., Govindarajan T.R., Ydri B., Fermion doubling
problem and noncommutative geomet\-ry~II,
\href{http://arxiv.org/abs/hep-th/0006216}{hep-th/0006216}.

\bibitem{fuzzyS} Grosse H., Reiter G., The fuzzy supersphere,
{\it J. Geom. Phys.}, 1998, V.28, 349--383,
\href{http://arxiv.org/abs/math-ph/9804013}{math-ph/9804013}.

\bibitem{klimcik1}
Klim\v{c}ik C.,  A nonperturbative regularization of the
supersymmetric Schwinger model, {\it Comm. Math. Phys.}, 1999,
V.206, 567--586,
\href{http://arxiv.org/abs/hep-th/9903112}{hep-th/9903112}.

\bibitem{klimcik1+} Klim\v{c}ik C., An extended fuzzy supersphere
and twisted chiral superf\/ields,  {\it Comm. Math. Phys.}, 1999,
V.206, 587--601,
\href{http://arxiv.org/abs/hep-th/9903202}{hep-th/9903202}.

\bibitem{seckin1}
Balachandran A.P., Kurkcuoglu S., Rojas E., The star product on
the fuzzy supersphere,  {\it JHEP}, 2002, N~7, Paper~056, 
21~pages, \href{http://arxiv.org/abs/hep-th/0204170}{hep-th/0204170}.

\bibitem{susybreaking} Balachandran A.P., Pinzul A., Qureshi B.,
SUSY anomalies break $N=2$ to $N=1$: the supersphere and the fuzzy
supersphere, {\it JHEP}, 2005, N~12, Paper~002, 14~pages,
\href{http://arxiv.org/abs/hep-th/0506037}{hep-th/0506037}.

\bibitem{seckin2}
Kurkcuoglu S.,  Non-linear sigma model on the fuzzy supersphere,
{\it JHEP}, 2004, N~3, Paper~062, 11 pages,
\href{http://arxiv.org/abs/hep-th/0311031}{hep-th/0311031}.



\bibitem{Szabo}
 Szabo R.J., D-branes in noncommutative f\/ield theory,
\href{http://arxiv.org/abs/hep-th/0512054}{hep-th/0512054}.

\bibitem{ezawa}
Ezawa Z.F., Tsitsishvili G., Hasebe K., Noncommutative geometry, extended $W_\infty$ 
algebra and Grassmannian solitons in multicomponent quantum Hall systems, {\it Phys. Rev.~B}, 2003,
V.67, 125314, 15~pages,
\href{http://arxiv.org/abs/hep-th/0209198}{\mbox{hep-th/0209198}}.


\bibitem{witten}
Seiberg N.,  Witten E., String theory and noncommutative geometry,
{\it JHEP}, 1999, N~9, Paper~032, 100~pages,
\href{http://arxiv.org/abs/hep-th/9908142}{hep-th/9908142}.

\bibitem{shiraz}
Minwala S., van Raamsdonk M., Seiberg N., Noncommutative
perturbative dynamics, {\it JHEP}, 2000, N~2, Paper~020, 33 pages,
\href{http://arxiv.org/abs/hep-th/9912072}{hep-th/9912072}.

\bibitem{Chaichian}
 Chaichian M., Kulish P.P., Nishijima K., Tureanu A., On a
Lorentz-invariant interpretation of noncommutative space-time and
its implications on noncommutative QFT, {\it Phys. Lett.~B}, 2004,
V.604, 98--102,
\href{http://arxiv.org/abs/hep-th/0408069}{hep-th/0408069}.

\bibitem{Chaichian1} Chaichian M., Presnajder P., Tureanu A., New concept of relativistic invariance in NC
space-time: twisted Poincar\'e symmetry and its implications, {\it
Phys. Rev. Lett.}, 2005, V.94, 151602, 15 pages,
\href{http://arxiv.org/abs/hep-th/0409096}{hep-th/0409096}.

\bibitem{Aschieri}
Dimitrijevic M., Wess J., Deformed bialgebra of dif\/feomorphisms,
\href{http://arxiv.org/abs/hep-th/0411224}{hep-th/0411224}.

\bibitem{Aschieri1} Aschieri P., Blohmann C., Dimitrijevic M.,
Meyer F., Schupp P., Wess J., A gravity theory on noncommutative
spaces, {\it Classical Quantum Gravity}, 2005, V.22, 3511--3532,
\href{http://arxiv.org/abs/hep-th/0504183}{hep-th/0504183}.

\bibitem{Drinfeld}
 Drinfel'd V.G., Quasi Hopf algebras, {\it Leningrad Math. J.}, 1990, V.1,
1419--1457.

\bibitem{Bal2}
 Balachandran A.P., Mangano G., Pinzul A., Vaidya S., Spin and
statistics on the Groenewold--Moyal plane: Pauli--Forbidden levels
and transitions, {\it Internat. J. Modern Phys.~A}, 2006, V.21,
3111--3126,
\href{http://arxiv.org/abs/hep-th/0508002}{\mbox{hep-th/0508002}}.

\bibitem{Bal2+} Qureshi B.A., Twisted supersymmetry,
fermion-boson mixing and removal of UV-IR mixing,
\href{http://arxiv.org/abs/hep-th/0602040}{\mbox{hep-th/0602040}}.




\bibitem{anca}
Tureanu A., Twist and spin-statistics relation in noncommutative
quantum f\/ield theory, {\it Phys. Lett.~B}, 2006, V.638,
296--301,
\href{http://arxiv.org/abs/hep-th/0603219}{hep-th/0603219}.

\bibitem{zahn}
Zahn J., Remarks on twisted noncommutative quantum f\/ield theory,
{\it Phys. Rev.~D}, 2006, V.73, 105005, 13~pages,
\href{http://arxiv.org/abs/hep-th/0603231}{hep-th/0603231}.

\bibitem{bald}
 Balachandran A.P., Govindarajan T.R., Mangano G., Pinzul A.,
 Qureshi B.A., Vaidya S., Statistics and UV-IR mixing with twisted
Poincar\'e invariance,
\href{http://arxiv.org/abs/hep-th/0608179}{hep-th/0608179}.

\bibitem{Majid}
 Majid S., Foundations of quantum group theory, Cambridge
University Press, 1995.

\bibitem{Oeckl}
 Oeckl R., Untwisting noncommutative $R^d$ and the equivalence of
quantum f\/ield theories, {\it Nuclear Phys. B}, 2000, V.581,
559--574,
\href{http://arxiv.org/abs/hep-th/0602040}{hep-th/0003018}.

\bibitem{Fiore}
 Fiore G., Schupp P., Statistics and quantum group symmetries,
in  Quantum Groups and Quantum Spaces, {\it Banach Centre
Publications}, Vol.~40, Warszawa, Institute of Mathematics, Polish
Academy of Sciences, 1997, 369--377.

\bibitem{Fiore1} Fiore G., Deforming maps and Lie
group covariant creation and annihilation operators, {\it J. Math.
Phys.}, 1998, V.39, 3437--3452,
\href{http://arxiv.org/abs/q-alg/9610005}{q-alg/9610005}.

\bibitem{Fiore2} Fiore G.,
Schupp P., Identical particles and quantum symmetries, {\it
Nuclear Phys. B}, 1996, V.470, 211--235,
\href{http://arxiv.org/abs/hep-th/9508047}{hep-th/9508047}.

\bibitem{Watts}
 Watts P., Noncommutative string theory, the $R$-matrix and Hopf
algebras, {\it Phys. Lett.~B}, 2000, V.474, 295--302,
\href{http://arxiv.org/abs/hep-th/9911026}{hep-th/9911026}.

\bibitem{Watts1} Watts P., Derivatives and the role of the
Drinfel'd twist in noncommutative string theory,
\href{http://arxiv.org/abs/hep-th/0003234}{hep-th/0003234}.

\bibitem{Bal3}
 Balachandran A.P., Pinzul A., Qureshi B.A., UV-IR mixing in
non-commutative plane, {\it Phys. Lett. B}, 2006, V.634, 434--436,
\href{http://arxiv.org/abs/hep-th/0508151}{hep-th/0508151}.

\bibitem{Bal4}
 Balachandran A.P., Pinzul A., Qureshi B.A., Vaidya S.,
Poincar\'e invariant gauge and gravity theories on the
Groenewold--Moyal plane,
\href{http://arxiv.org/abs/hep-th/0608138}{hep-th/0608138}.




\end{thebibliography}
\end{document}